\documentclass{article}
\usepackage{graphicx}
\usepackage{color}
\definecolor{ForestGreen}{rgb}{0.13, 0.55, 0.13}
\definecolor{airforceblue}{rgb}{0.36, 0.54, 0.66}
\definecolor{orange}{rgb}{1.0, 0.5, 0.0}

\newcommand{\fnl}{f_{\rm NL}^{\rm loc}}
\newcommand{\bx}{\mathbf{x}}

     \usepackage[final,nonatbib]{neurips_2020_tda}

\usepackage[utf8]{inputenc} 
\usepackage[T1]{fontenc}    
\usepackage{amsfonts}       
\usepackage{booktabs}       
\usepackage{hyperref}       
\usepackage{url}            
\usepackage{microtype}      
\usepackage{nicefrac}       
\usepackage{paralist}       
\usepackage{siunitx}        
\usepackage{amsmath}
\usepackage{color}

\title{
Topological Echoes of Primordial Physics\\ in the Universe at Large Scales
}

\author{%
    Alex Cole \\
  Gravitation Astroparticle Physics Amsterdam (GRAPPA) \\
  Institute for Theoretical Physics Amsterdam \\
  University of Amsterdam, the Netherlands \\
   \texttt{a.e.cole@uva.nl} \\
   \AND
  Matteo Biagetti \\
  IFPU Trieste, Italy \\
  \texttt{matteo.biagetti@sissa.it} \\
   \And
   Gary Shiu \\
   Department of Physics\\ University of Wisconsin-Madison, USA \\
   \texttt{shiu@physics.wisc.edu} \\
}

\begin{document}

\maketitle

\begin{abstract}
 We present a pipeline for characterizing and constraining initial conditions in cosmology via persistent homology. The cosmological observable of interest is the cosmic web of large scale structure, and the initial conditions in question are non-Gaussianities (NG) of primordial density perturbations. We compute persistence diagrams and derived statistics for simulations of dark matter halos with Gaussian and non-Gaussian initial conditions. For computational reasons and to make contact with experimental observations, our pipeline computes persistence in sub-boxes of full simulations and simulations are subsampled to uniform halo number. We use simulations with large NG ($\fnl=250$) as templates for identifying data with mild NG ($\fnl=10$), and running the pipeline on several cubic volumes of size $40~(\si{Gpc/h})^{3}$, we detect $\fnl=10$ at $97.5\%$ confidence on $\sim 85\%$ of the volumes for our best single statistic. Throughout we benefit from the interpretability of topological features as input for statistical inference, which allows us to make contact with previous first-principles calculations and make new predictions.
\end{abstract}

\section{Introduction}
The last few decades have established a concordance model of cosmology, which relies on a handful of very well-constrained parameters \cite{Aghanim:2018eyx,Alam:2020sor}. Despite this progress, mysteries remain. Perhaps the most pressing question is the mechanism that seeded primordial perturbations. These perturbations evolved under gravity to form the cosmic structure we observe today. By studying the statistics of cosmological observables, we aim to reverse-engineer the theory that generated primordial perturbations. Observations of the cosmic microwave background have so far provided the strongest constraints on the two-point function of primordial perturbations \cite{Akrami:2018odb}, which  however only provides coarse-grained information about the free theory of the fields generating these perturbations, while not answering fundamental questions about the field content and interactions during this early phase. To probe the precise microphysics of this theory, we must move beyond the two-point function. One way to do this is to search for evidence of primordial non-Gaussianity (NG), the precise form of which allows one to discriminate between different theories of the very early universe \cite{Bartolo:2004if}. For example, local NG \cite{Komatsu:2001rj} is described by the ansatz
\begin{equation}
    \Phi_{\rm NG}(\bx)=\Phi_{\rm G}(\bx)+f_{\rm NL}^{\rm loc}\left(\Phi_{\rm G}^2(\bx)-\left<\Phi_{\rm G}^2(\bx)\right>\right)^2
\end{equation}
where $\Phi_{\rm NG}$ is the primordial gravitational potential, $\Phi_{\rm G}$ is a Gaussian random field, and $\fnl$ controls the amount of NG. Single field models necessarily produce very small local NG  as implied by consistency relations in the squeezed limit \cite{Maldacena:2002vr,Creminelli:2004yq}, though large NG can be realized in other shapes \cite{Chen:2006nt}.
Primordial NG has still not been detected in any form. Of the various forms of NG, local NG is the most tightly constrained, with $\fnl=-0.9\pm 10.2$ at $2\sigma$ \cite{Akrami:2019izv}.

There are several reasons to expect that topology is sensitive to NG. Namely, departures from Gaussianity are expected via an Edgeworth expansion to be most noticeable in the tail of a probability distribution. In terms of the primordial density perturbation, these tails are probed by local minima and local maxima. Under gravity, local maxima (overdense regions) collapse to form dark matter halos while local minima (underdense regions) expand as cosmic voids. These objects, along with filament loops, form a topological decomposition of large-scale structures (LSS) in the universe. 

We therefore propose to robustly compute the topology of LSS and corresponding constrain cosmology using persistent homology \cite{edelsbrunner2010computational,zomorodian2005topology,oudot2015persistence}. In the context of LSS, persistent homology tracks the creation and destruction of clusters, filament loops, and voids across scales. While certain qualitative aspects of persistent homology have been computed before for LSS \cite{vandeWeygaert:2011ip,vandeWeygaert:2013kma,Pranav:2016gwr}, the ability to perform statistics and generate constraints is underexplored. As the first work in this vein, we take into account several important observational effects, including varying halo density. Additionally, this is the first computation of the persistent topology of LSS with non-Gaussian initial conditions.
In a cosmology context, the statistics derived from persistent homology have the advantage over other ML techniques of interpretability: we are able to reproduce previous results in certain limits and we make new predictions for unexplored observables, such as filament loops formed by dark matter halos.

This is a condensed version of the paper \cite{Biagetti:2020skr}.
\section{Method}
\subsection*{Data set, subboxing, subsampling}
We use the \textsc{Eos Dataset}\footnote{Information on the full dataset is available at \href{https://mbiagetti.gitlab.io/cosmos/nbody/eos/}{https://mbiagetti.gitlab.io/cosmos/nbody/eos/}. The original paper investigated the imprint of primordial non-Gaussianity on the halo power spectrum \cite{Biagetti:2016ywx}.}, a suite of N-body simulations created to investigate the imprint of primordial non-Gaussianity in LSS at low redshift. The nomenclature for the cosmologies in the dataset is as follows: \textsf{(G/NG)\#(L/S)}, where \textsf{G/NG} distinguishes Gaussian from non-Gaussian initial conditions, \textsf{\#} gives the value of the cosmological parameter $\sigma_8$ in the \textsf{G} case or the value of $\fnl\neq 0$ in the \textsf{NG} case (\textsf{NG} simulations have $\sigma_8=0.85$, the same $\sigma_8$ value as in \textsf{G85} simulations), and $\textsf{L/S}$ determines whether a simulation was computed in a ``large box'' of side length of $2~\si{Gpc/h}$ or a ``small box'' of side length of $1~\si{Gpc/h}$ with different resolution. Each simulation in \textsf{L} has $\mathcal{O}(10^6)$ DM halos at redshift $z=1$. When performing statistics, we will consider concatenated simulations with volume $40~(\si{Gpc/h})^3$. In the end we are considering similar volumes, redshift ranges, and DM halo mass ranges as those measured by upcoming galaxy surveys like Euclid. One difference to keep in mind is that galaxy surveys observe luminous galaxies hosted by DM halos, while restricting to DM-only simulations allows us to maintain control over the relevant physics.

The size of individual realizations in the \textsc{Eos Dataset} presents a challenge for persistent homology. To deal with it, we use filtrations built from subcomplexes of the Delaunay triangulation, as well as computing filtrations and their persistent homology in \emph{sub-boxes} of a full simulation with side length $1~\si{Gpc/h}$. This is possible without introducing systematics because the topological features in our filtrations die at scales of $\mathcal{O}(10)\si{Mpc/h}\ll 1~\si{Gpc/h}$.

An additional complication in the context of experimental observations of LSS is cosmic variance, which means that the number of halos in a fixed simulation volume can vary between simulations. This effect can compete with variations in halo density due to cosmology, and can have a large effect on persistence diagrams. To remove this systematic, we \emph{subsample} each simulation to have the same number of halos. Without subsampling, two cosmologies that increase the total number of halos are much more difficult to distinguish via persistent homology, while subsampled simulations can be distinguished \cite{Biagetti:2020skr}.
\subsection*{Filtrations and statistics}
Our primary filtration is a modification of the DTM-based filtration of \cite{2018arXiv181104757A}. In particular, we consider subcomplexes of the Delaunay triangulation organized by both scale and the extent to which a particular point is an outlier. We call this an $\alpha$DTM$\ell$-filtration. The outlier property of a point is quantified by the Distance-to-Measure (DTM) function. Given a set of points $X$, a positive integer $k$, and a positive real number $p$, the empirical DTM function is given by \cite{chazal2011geometric}

\begin{equation}
    {\rm DTM}(x)=\frac{1}{k}\left(\sum_{X_i\in N_k(x)}||x-X_i||^p\right)^{1/p}
\end{equation}
where $N_k(x)$ is the list of the $k$ nearest neighbors in $X$ to $x$. We will take $p=2$ and $k=15$.
Now, given a point $x\in X$ and a positive real number $q$ we consider the radius function
\begin{equation}
    r_x(\nu)=\left(\nu^q-{\rm DTM}(x)^q\right)^{1/q}\quad \textrm{if }\nu\geq {\rm DTM}(x)
\end{equation}
and $-\infty$ otherwise, and by definition a ball of radius $r=-\infty$ is the empty set. The role of ${\rm DTM}(x)$ is to delay the contribution of outliers to the filtration. 
Here $q$ controls the mixing between ${\rm DTM}(x)$ and $\nu$. We will take $q=2$.  For very large $q$, $r_x(\nu)$ is $-\infty$ until $\nu={\rm DTM}(x)$, at which point it quickly jumps to $\nu$. Beyond the vertices, we add an edge $[x_1 x_2]$ that is present in the Delaunay complex once the relevant balls overlap, i.e.\ $B_{r_{x_1}(\nu)}\cap B_{r_{x_2}(\nu)}\neq \emptyset$. Triangles and tetrahedra are then added when all necessary faces are present. In this construction, topological features will suffer less from the presence of outliers that in the $\alpha$-filtration.

From the persistence diagrams corresponding to these filtrations, we derive several summary statistics for the purpose of inference. These include Persistence Images (PIs) \cite{adams2017persistence} and the ``topological curves'' defined by
\begin{align}
    B_p(\nu)&=\textrm{count}_p(\nu>\nu_{\rm birth})\\
    D_p(\nu)&=\textrm{count}_p(\nu>\nu_{\rm death})\\
    P_p(\nu)&=\textrm{count}_p(\nu>\nu_{\rm persist})\\
    b_p(\nu)&=D_p(\nu)-B_p(\nu)
\end{align}
The curves $B_p,D_p,P_p$ correspond up to normalization to empirical distribution functions of births, deaths, and persistences of $p$-cycles in a given diagram, while $b_p$ are the familiar Betti numbers. {Note that once a binning in $\nu$ is specified, these curves live in finite-dimensional vector spaces.}
\section{Results}
\begin{figure}
\centering
\includegraphics[width=0.33\textwidth]{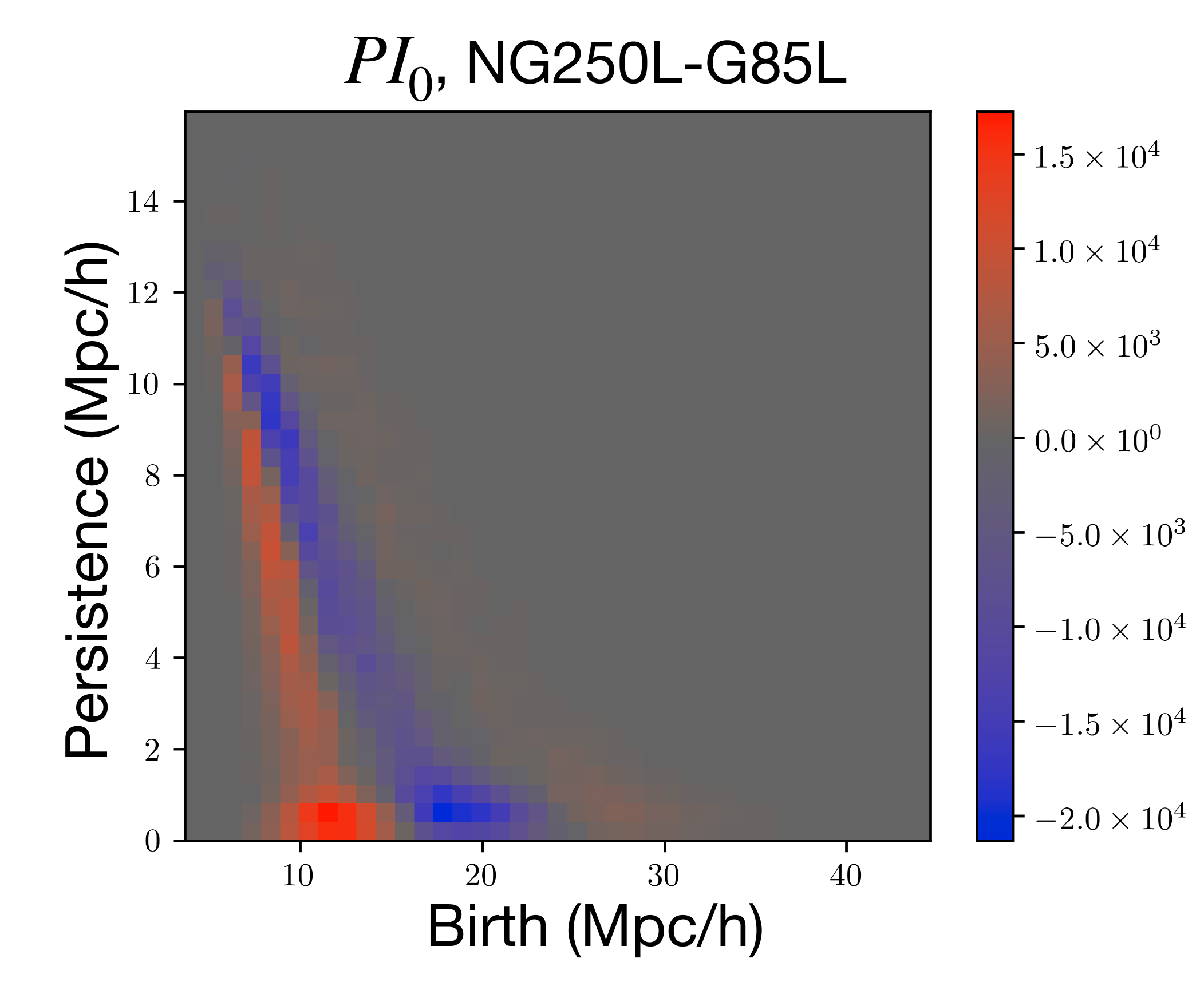}\includegraphics[width=0.33\textwidth]{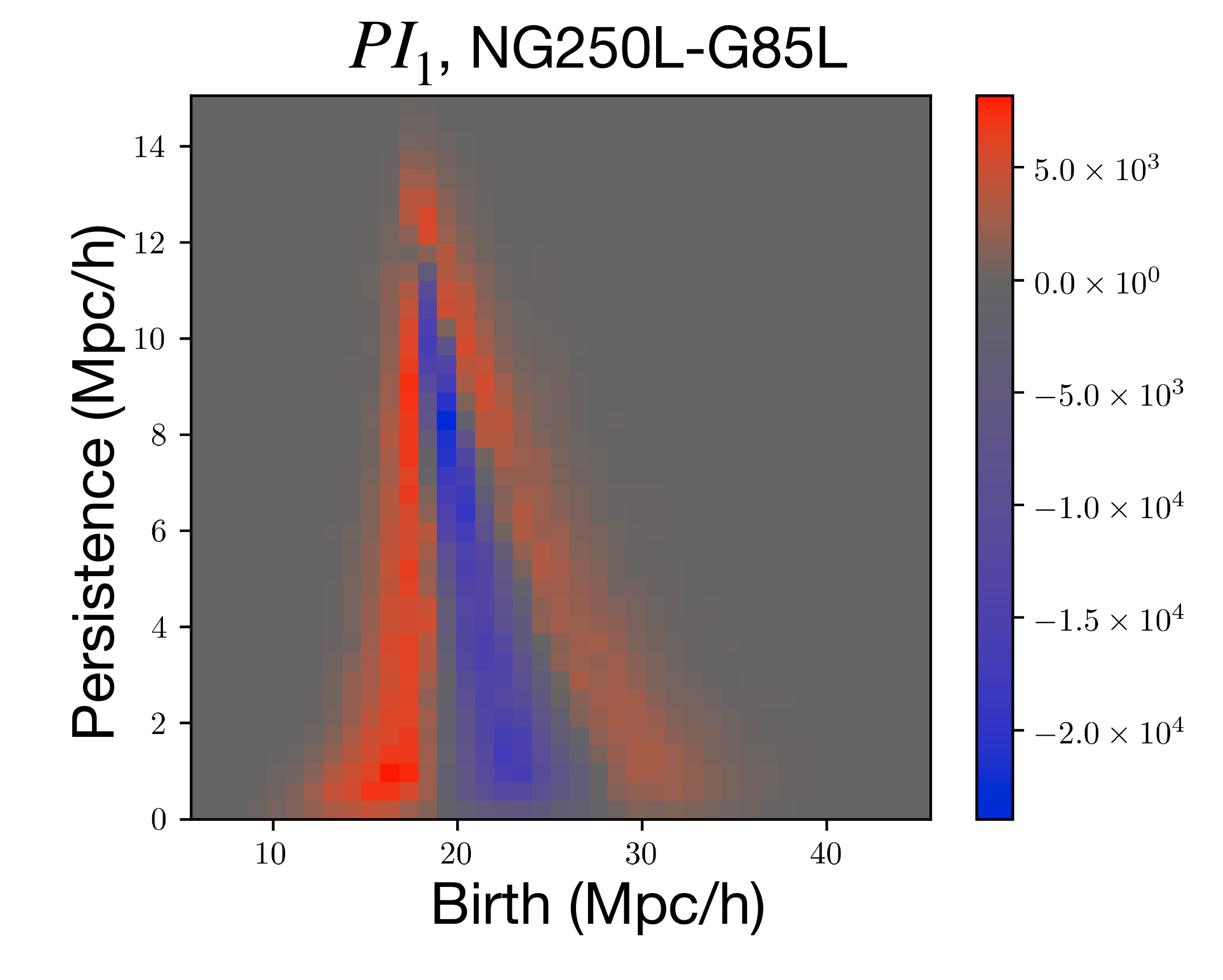}\includegraphics[width=0.33\textwidth]{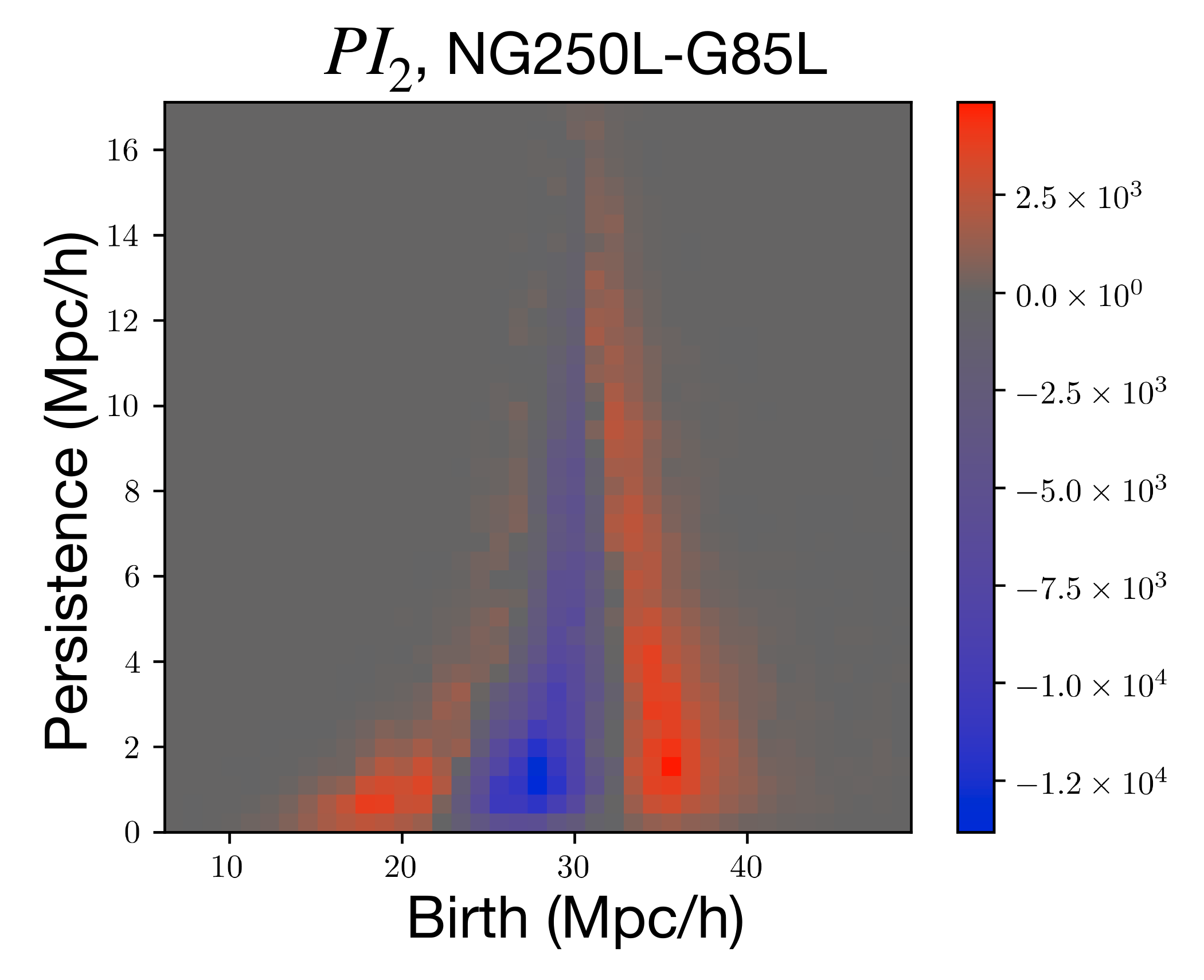}
\includegraphics[width=0.33\textwidth]{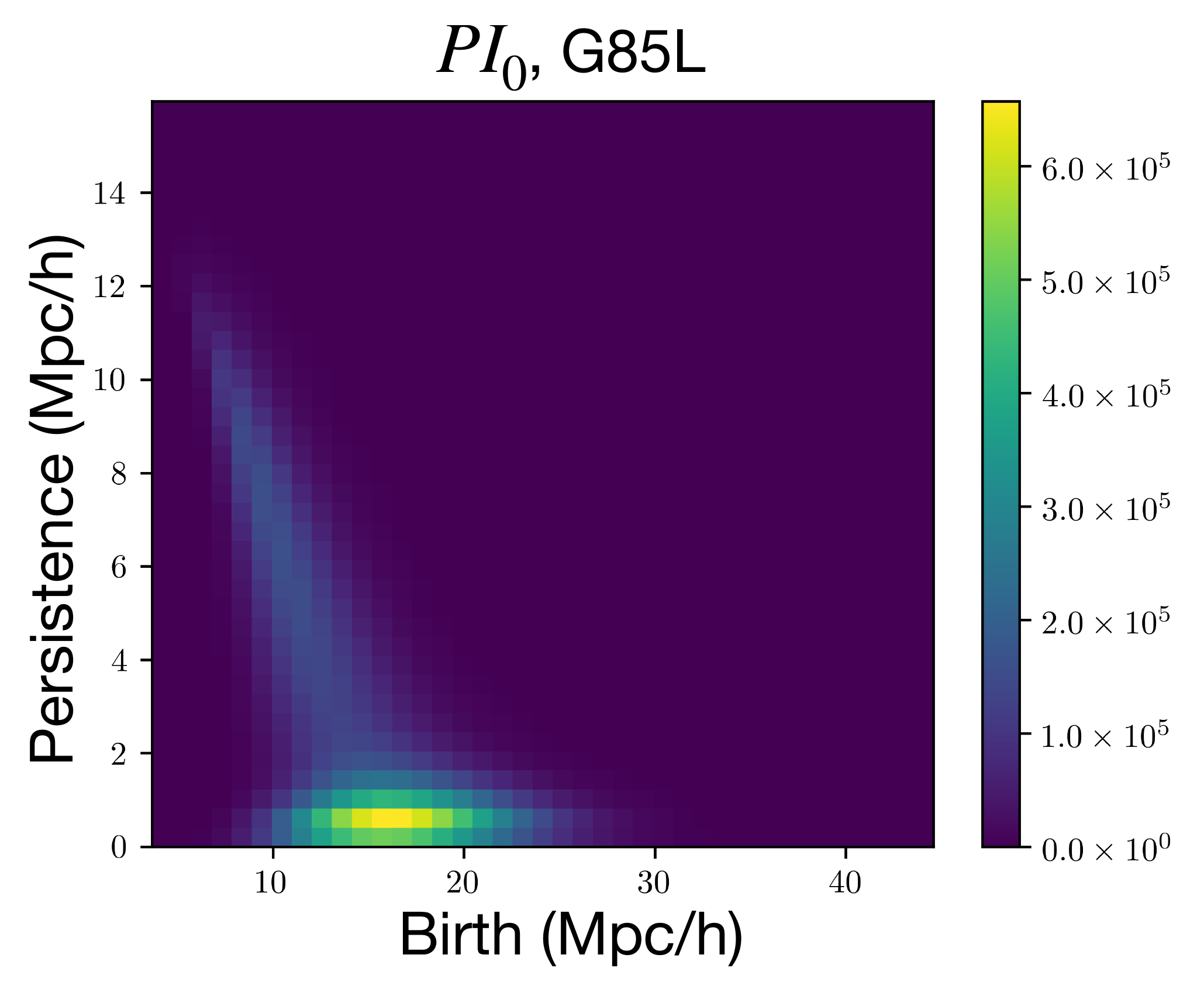}\includegraphics[width=0.33\textwidth]{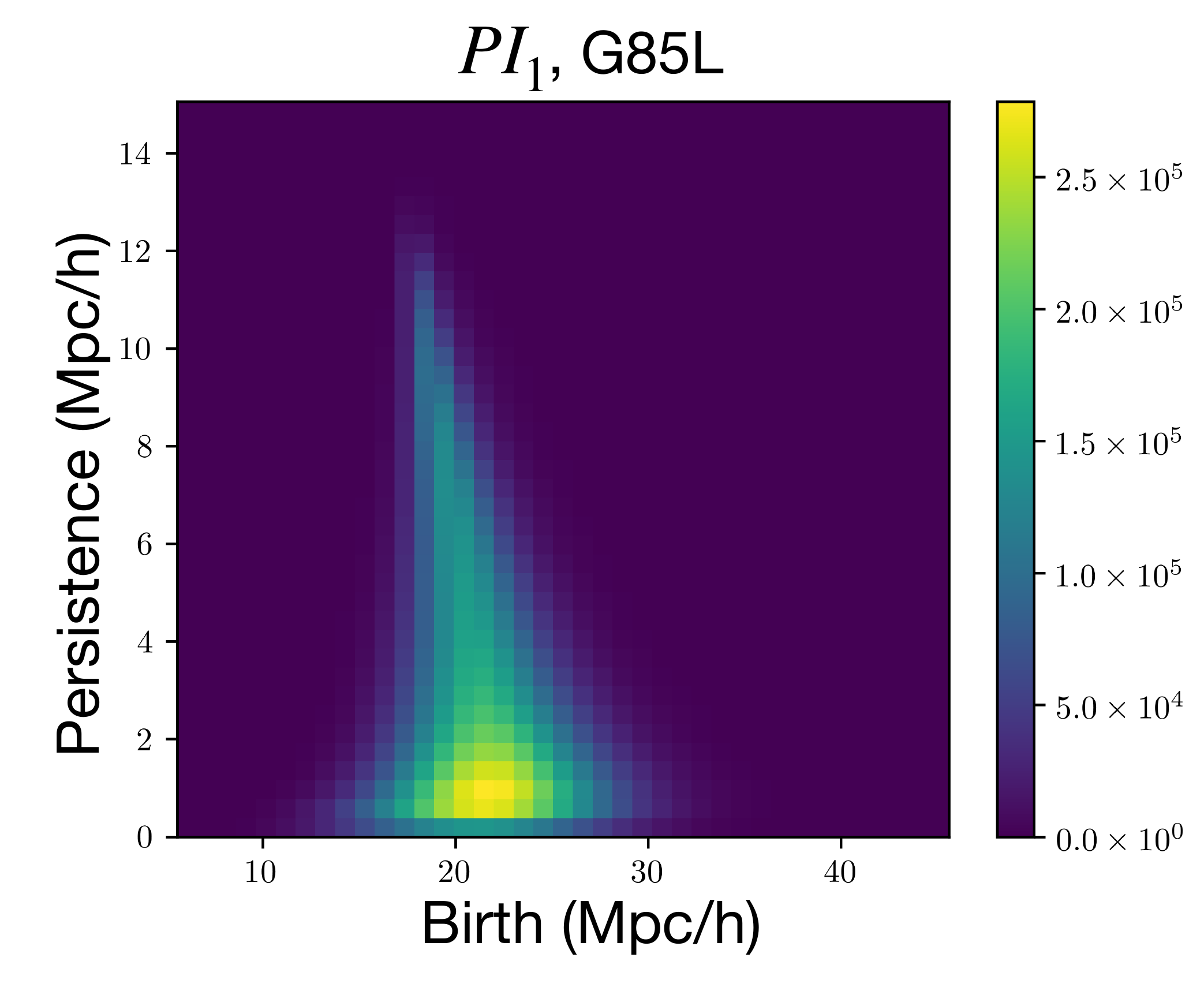}\includegraphics[width=0.33\textwidth]{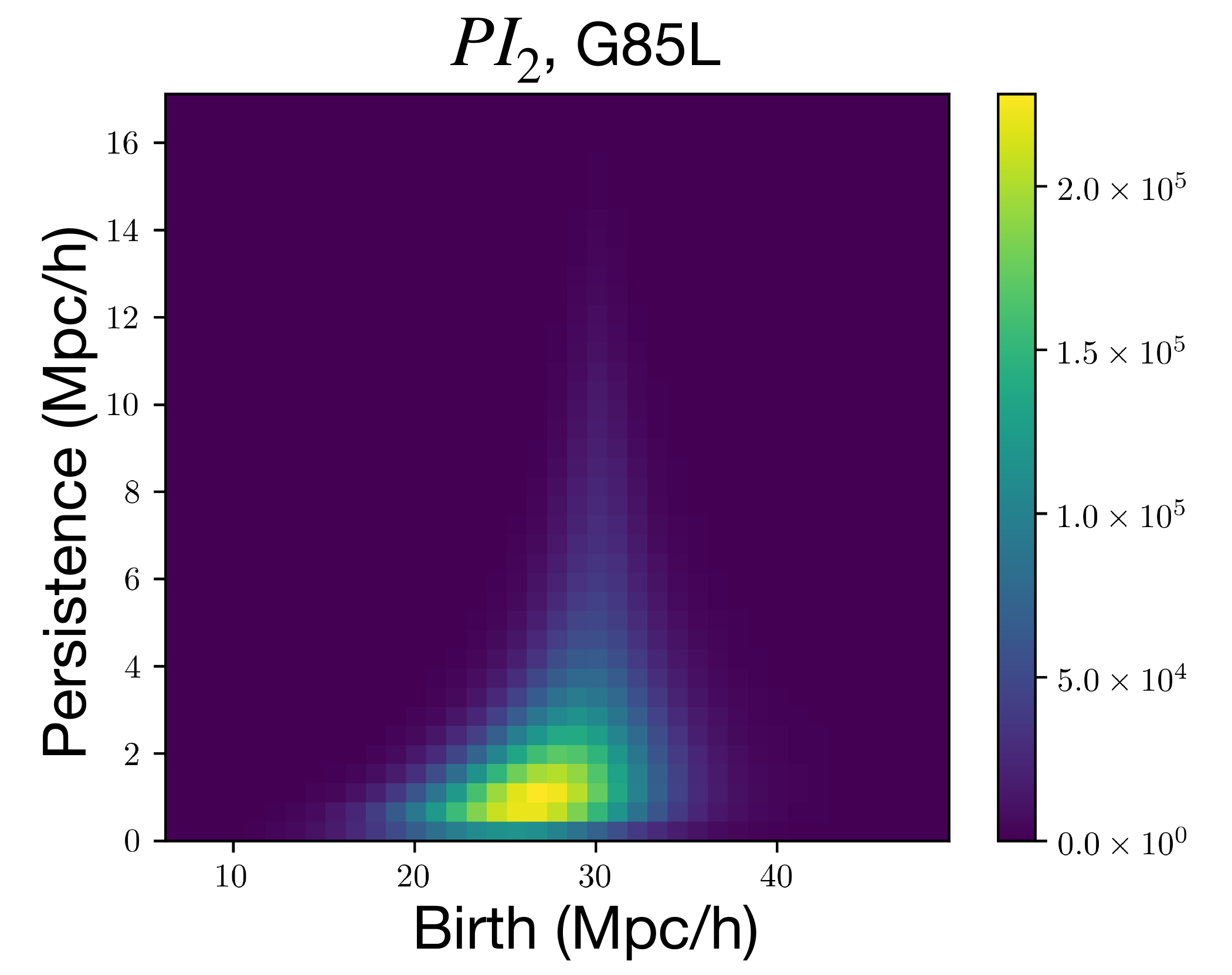}
\caption{The effect of local non-Gaussianity on PIs for the $\alpha$DTM$\ell$-filtration. In the top row the difference between the PIs averaged over \textsf{NG250L} and \textsf{G85L} is shown. The bottom row shows the PIs averaged over \textsf{G85L}. 
}\label{fig:PIintuition}
\end{figure}
We show the average PIs and their differences for our simulations in Figure \ref{fig:PIintuition}. We examine the signature generated by $\fnl=250$, 
as the signal is large and results in clean patterns when averaged over our simulation volume. In the end we are interested in ``detecting'' $\fnl=10$ using these patterns. The effects of a nonzero $\fnl$ on all the topological features share a common turnover feature: non-Gaussian simulations dominate the statistics for small and large birth scales while at intermediate scales, Gaussian simulations dominate. The PIs suggest that the topological curves $B_p(\nu)$, $D_p(\nu)$, and $b_p (\nu)$  should outperform $P_p(\nu)$ in detecting $\fnl$.

{
Recall that 0-cycles in our filtrations are born at the DTM values of their respective halos. Examining $PI_0$, we see that $\fnl>0$ mainly causes a population of 0-cycles to shift to earlier births, i.e.\ smaller DTM values, while a subdominant population shifts to larger DTM values. This is consistent with the heuristic observation that positive $\fnl$ affects structure as if there were more gravitational evolution in overdense regions (lowering the DTM values of corresponding halos as the neighborhood becomes more dense with halos) and less in underdense regions (increasing DTM values as fewer halos are formed there) \cite{Dalal:2007cu,Kang:2007gs,Grossi:2008fm,Pillepich:2008ka}. }
\begin{figure}
    \centering
    \includegraphics[width=0.33\textwidth]{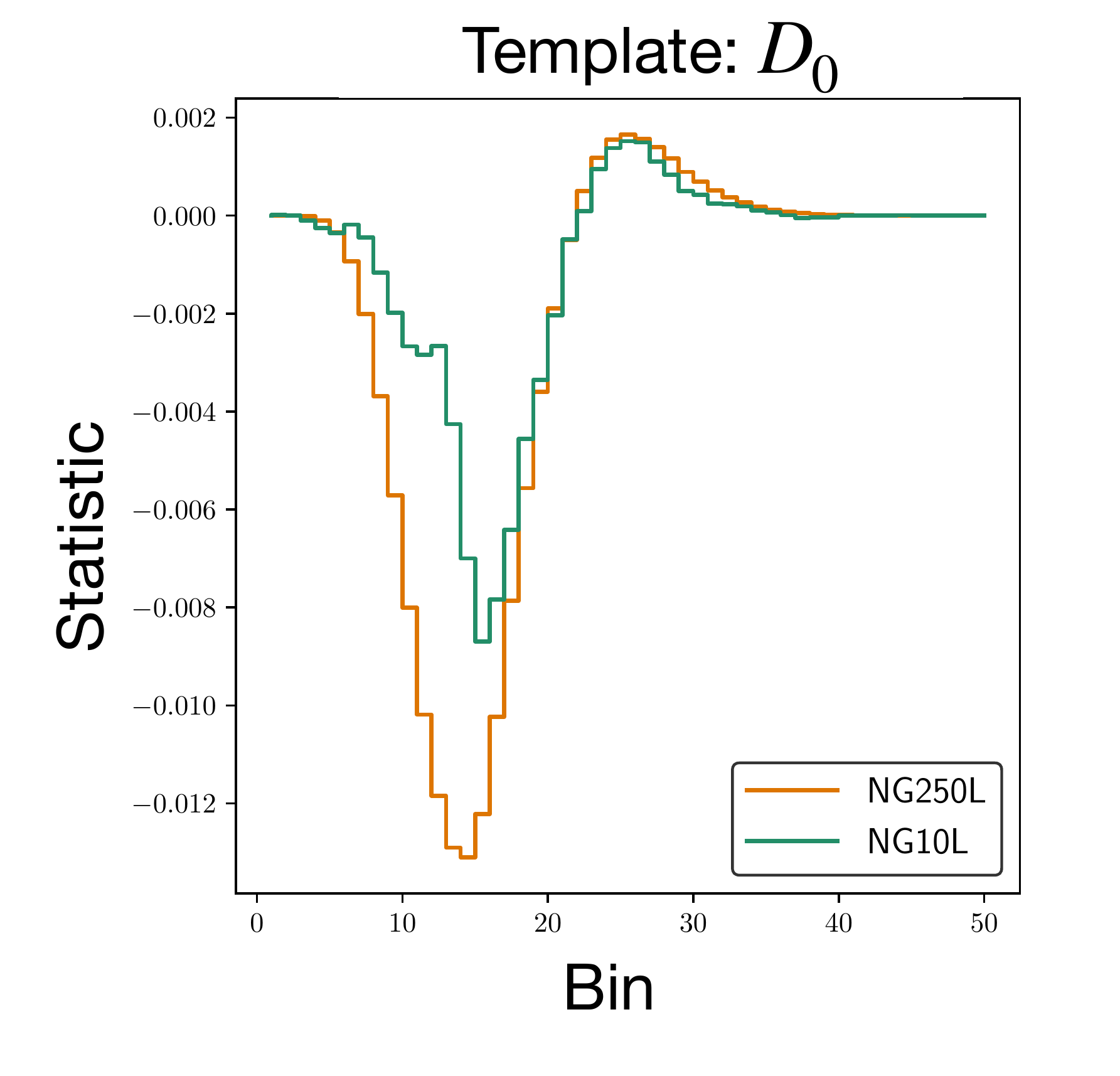}\includegraphics[width=0.31\textwidth]{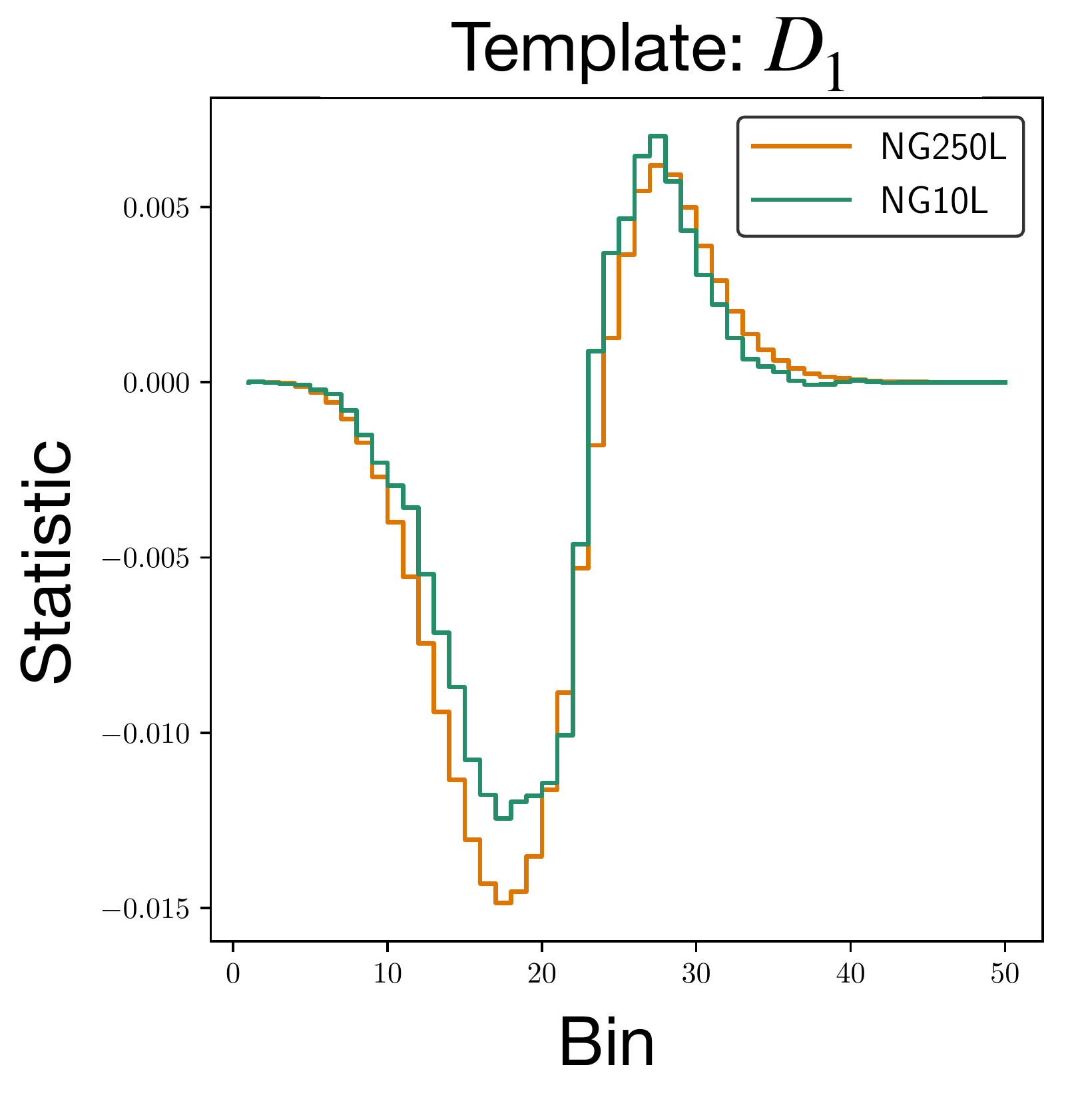}\includegraphics[width=0.33\textwidth]{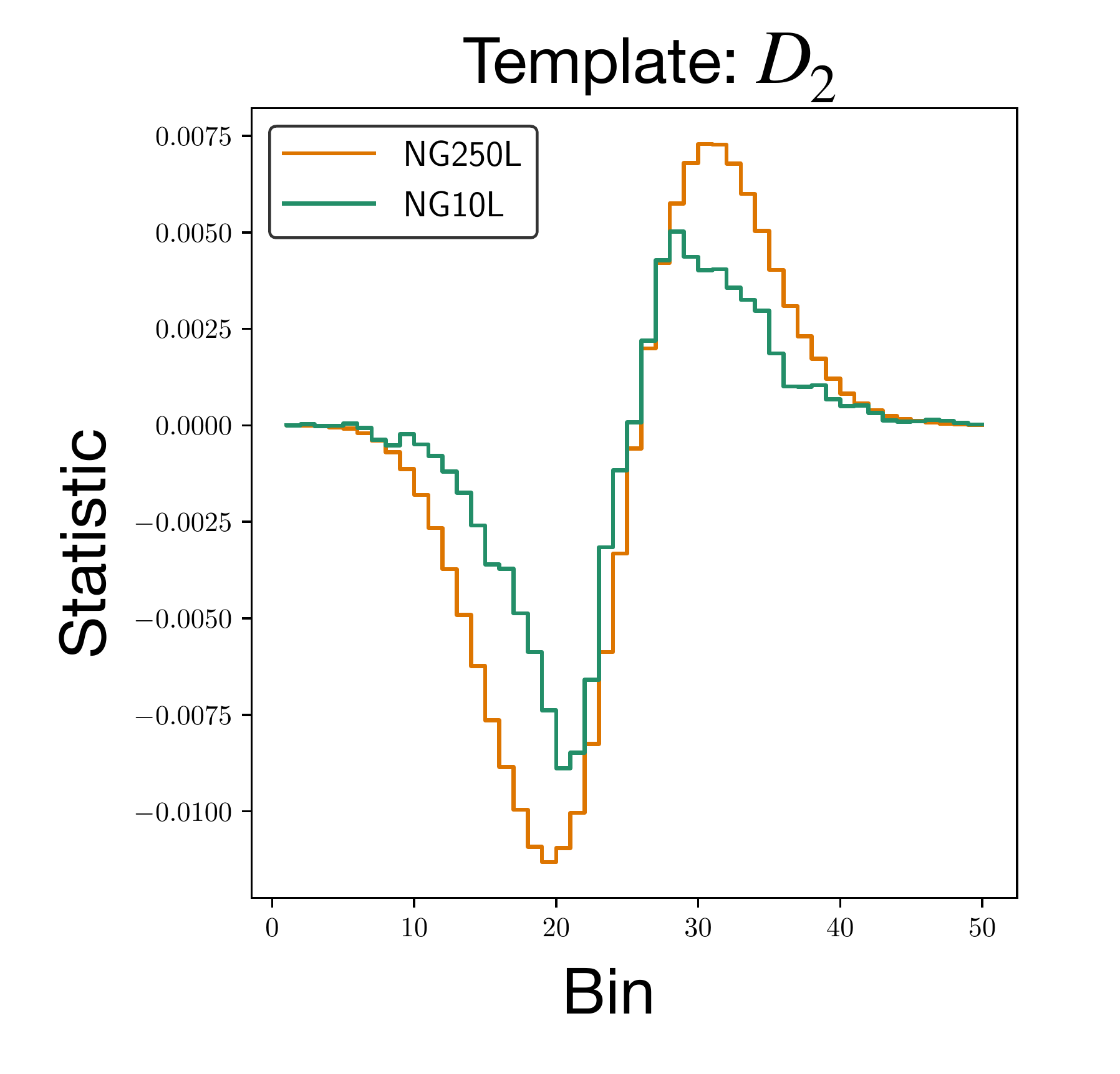}
    \caption{The effect of $\fnl>0$ on $D_p$. We show $D_p(\fnl>0)-D_p(\fnl=0)$ averaged over realizations with non-Gaussian simulations taken either from \textsf{NG10L} or \textsf{NG250L} and Gaussian simulations taken from \textsf{G85L}. 
    }
    \label{fig:deathDists}
\end{figure}

{ Among the topological curves we consider, $D_p$ is correlated with the sizes of topological features. We show the effect of $\fnl>0$ on $D_p$ in Fig.\ \ref{fig:deathDists}. Examining these more closely, we find that $D_2$ exhibits features consistent with previously derived results \cite{Kamionkowski:2008sr}. Ref.\ \cite{Kamionkowski:2008sr} performed a first-principles calculation of the effect of $\fnl$ of the distribution of void sizes, finding that $\fnl$ would increase the number of voids of small radius and decrease the number of voids of large radius. The manifests as the initial dip and subsequent rise of $D_2(\fnl>0)-D_2(\fnl=0)$ in Fig.\ \ref{fig:deathDists}. At large scales, we again measure a larger fraction of voids dying in the non-Gaussian case, a prediction not made in the approximation scheme of \cite{Kamionkowski:2008sr}.

Intriguingly, a similar structure in $D_1$ suggests that similar physics is at play for filament loops. However, for these observables first-principles results are lacking, in part because filament loops break spherical symmetry. Our results motivate the development of analytical techniques like those in \cite{Kamionkowski:2008sr} for studying filament loops.}

All in all, the interpretability of our statistics allows us to make contact with previously derived quantities, while motivating future study to derive other aspects of our results from first principles.

We can perform the following statistical test: for a statistic $\vec{S}$ (e.g. $B_1$, here written as a vector), we compute the \emph{template} $\vec{T}_{\rm NG}\equiv\vec{S}_{\textsf{NG250S}}-\vec{S}_{\textsf{G85S}}$, where the subscript denotes averaging over all realizations. Given a simulation to test for NG, we compute the topological statistic $\vec{S}_{\rm test}$ and then compute the distance
\begin{equation}
    D_{\rm template}=\frac{\vec{S}_{\rm test}\cdot \vec{T}_{\rm NG}-\vec{S}_{\textsf{G85L}}\cdot \vec{T}_{\rm NG}}{\sigma}
\end{equation}
where $\sigma$ describes the variance of $\vec{S}\cdot\vec{T}_{\rm NG}$ within \textsf{G85L}. { Intuitively, if the test simulation has NG initial conditions, $D_{\rm template}$ will be positive via the construction of $\vec{T}_{\rm NG}$.
To generate quantitative predictions, we set a baseline by using test simulations drawn from \textsf{G85L}. This case models our uncertainty due to cosmic variance: even if some actual galaxy survey data had the exact same cosmological parameters as \textsf{G85L}, it would still be seeded by a different realization of a random field. }

\begin{figure}[h]
    \centering
    \includegraphics[width=0.45\textwidth]{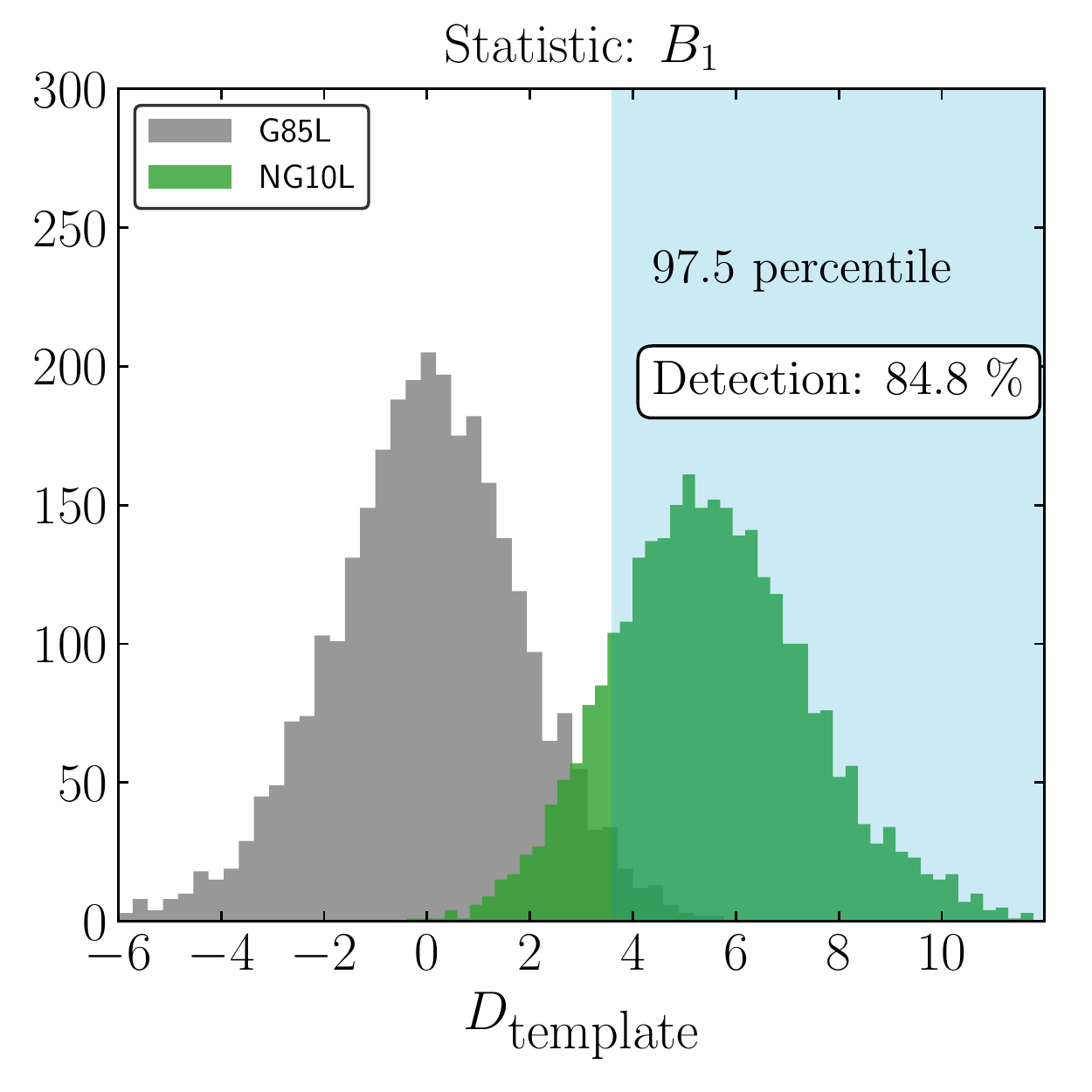}
    \includegraphics[width=0.45\textwidth]{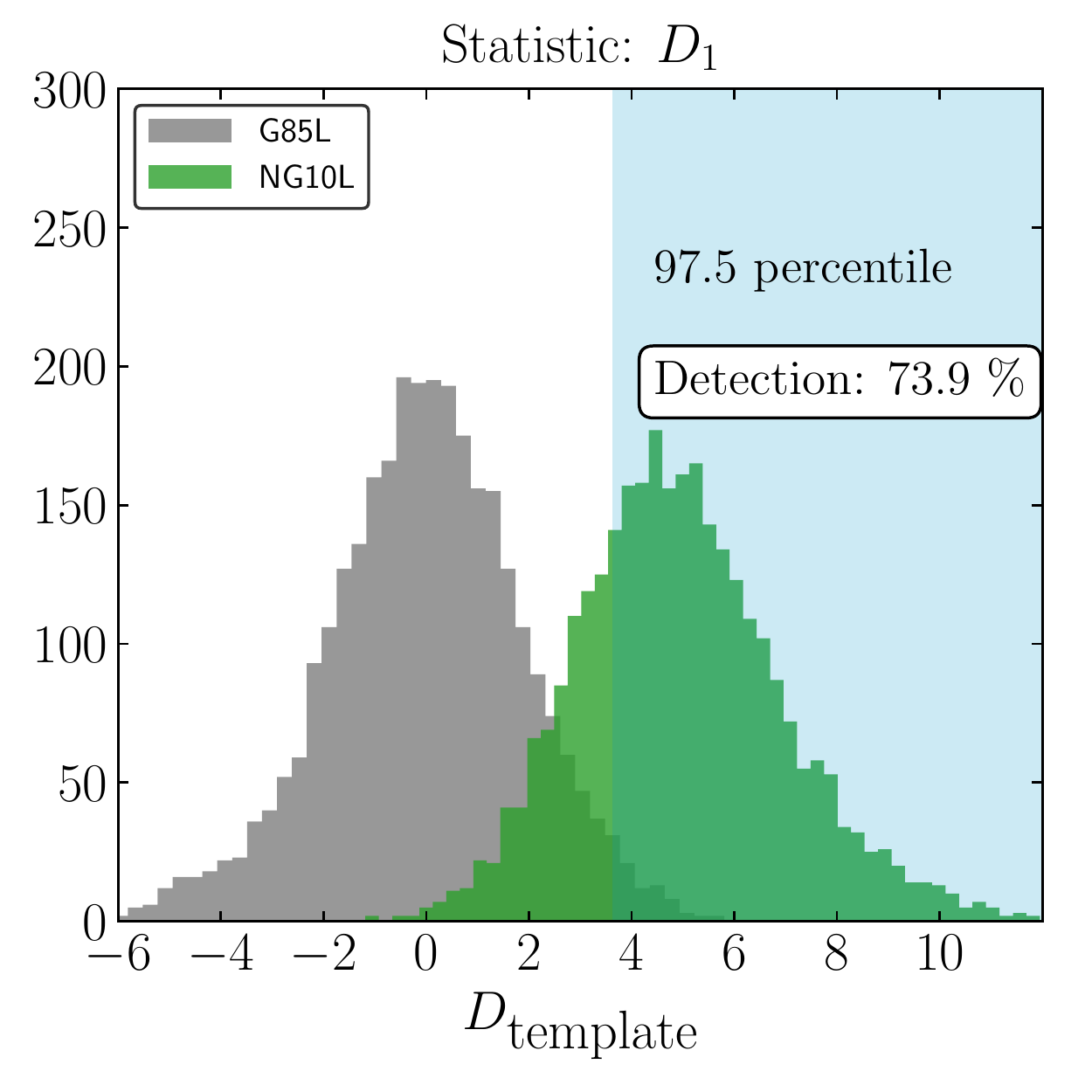}
    \caption{Distribution of $D_{\rm template}$. Each data point represents a volume of $40~(\si{Gpc/h})^3$.  The fraction of \textsf{NG10L} data points lying beyond the 97.5\% threshold of the \textsf{G85L} distribution is quoted as the detection rate of $\fnl=10$.}
    \label{fig:TemplateHist}
\end{figure}

We run this pipeline with test simulations taken from \textsf{G85L} and for \textsf{NG10L} (being careful to use distinct simulation seeds). We use the distribution of $D_{\rm template}$ when the test data comes from \textsf{G85L} to set a $97.5\%$ detection threshold for \textsf{NG10L}, see Fig.\ \ref{fig:TemplateHist}.
Each data point in Fig. \ref{fig:TemplateHist} reoresents a combination of simulations with total volume $40~(\si{Gpc/h})^{3}$. We find that for the statistic $B_1$, NG at a level $\fnl=10$ is detected at $97.5\%$ confidence for $\sim 85\%$ of these volumes.

{The standard approach to detecting NG in LSS relies on the two- and three-point correlations of galaxies (or in our case DM halos). Recently, a rigorous study of these correlation functions for the \textsc{Eos Dataset} was not able to detect $\fnl=10$ at $2\sigma$ \cite{MoradinezhadDizgah:2020whw}. Our current topological methods are sensitive to higher-order correlations in the data, allowing them to probe more subtle patterns in the distribution of halos.}

\section{Discussion}
We have introduced several practical considerations for persistent homology including subsampling and the use of sub-boxes to generate quantitative constraints on cosmology within a set of simulations.
While this pipeline was presented in terms of detecting primordial non-Gaussianity, it is applicable to constraining any cosmological parameter. In future work we plan to study the effect of a broad range of cosmological parameters on these topological observables. We plan to eventually apply this pipeline to galaxy survey data, for which systematics like baryon feedback and redshift space distortions must be taken into account.

Our code, which proves a convenient API for loading halo catalogs and computing their persistent topology, summary statistics, and corresponding constraints, is provided publicly.\footnote{The code is available at \href{https://gitlab.com/mbiagetti/persistent_homology_lss}{https://gitlab.com/mbiagetti/persistent\_homology\_lss}.}
\clearpage

\bibliographystyle{utphys}
\bibliography{Halos-TDA}

\end{document}